\documentclass[10pt,twocolumn,floatfix,footinbib]{revtex4-1}
\pdfoutput=1
\usepackage{natbib}\usepackage[colorlinks=true,citecolor=blue,linkcolor=blue,breaklinks=true]{hyperref}
\usepackage{amsmath,amssymb,mathtools}
\usepackage{epsfig}  
\usepackage{graphicx}   
\usepackage{slashed}             
\usepackage{url}
\usepackage{color}
\usepackage{multirow}
\usepackage{placeins}
\usepackage[dvipsnames]{xcolor}
\usepackage{lipsum}

\allowdisplaybreaks

\bibpunct{[}{]}{,}{n}{}{,}

\newcommand{\ba}{\begin{array}}
\newcommand{\ea}{\end{array}}
\newcommand{\bd}{\begin{displaymath}}
\newcommand{\ed}{\end{displaymath}}
\newcommand{\beq}{\begin{equation}}
\newcommand{\eeq}{\end{equation}}
\newcommand{\bea}{\begin{eqnarray}}
\newcommand{\eea}{\end{eqnarray}}

\newcommand{\ra}{\rightarrow}
\newcommand{\nn}{\nonumber}

\def\etc{ {\em etc.\ }}
\def\ie{ {\em i.e.,\ }}

\def\g{\gamma}

\def\m{\mu}
\def\n{\nu}

\def\q2 {q^2}

\def\bt{\begin{table}}
\def\et{\end{table}}

\catcode`@=11 
\def \gsim{\mathrel{\mathpalette\@versim>}}
\def \lsim{\mathrel{\mathpalette\@versim<}}
\def \@versim#1#2{\lower0.4ex\vbox{\baselineskip\z@skip\lineskip\z@skip
     \lineskiplimit\z@\ialign{$\m@th#1\hfil##\hfil$%
     \crcr#2\crcr\sim\crcr}}}
\catcode`@=12 

\begin{document}

\title{XENON1T constraints on neutrino non-standard interactions}

\author{Siddhartha Karmakar}
\email{ phd1401251010@iiti.ac.in}
\author{Sujata Pandey}
\email{phd1501151007@iiti.ac.in}
\affiliation{\em Discipline of Physics, Indian Institute of Technology Indore,\\
 Khandwa Road, Simrol, Indore - 453\,552, India}
 

\begin{abstract}
\noindent The new XENON1T observation of dark matter-electron scattering cross-section, along with further constraining many popular dark matter models, has indicated the possibility of new physics at a low energy. We point out that this new observation also significantly constrain the neutrino non-standard interactions~(NSI). We consider the NSI arising from a kinetically mixed $Z'$ with renormalisable and dipole-like interactions with the active and light sterile neutrinos. In passing, we also address the possibility of explaining the XENON1T excess around electron recoil energy $\sim 2$~keV in presence of such NSI.  
\end{abstract}

\maketitle

\section{Introduction}
\noindent The direct detection~(DD) experiments are one of the key ways to search for dark matter~(DM) owing to the weakly-interacting massive particle~(WIMP) paradigm. These experiments are designed to observe nuclear recoil in $\sim 1 - 100$~keV range while using noble gases or certain dense crystals as the medium. DM particles in our galactic halo can potentially scatter off the nuclei in these detectors, the non-observation of which puts constraints on DM-nucleon scattering cross-section. 
XENON1T~\cite{Aprile:2017aty} recently updated the constraints on  DM-electron scattering cross-section~\cite{Aprile:2020tmw}, which makes it the most restrictive bound of its kind at this point.  
Also, the XENON1T collaboration reported an excess around electron recoiling energy $\sim 2$~keV   in the detector~\cite{Aprile:2020tmw}. Numerous attempts to interpret this excess have been made in the context of dark matter~\cite{DelleRose:2020pbh, An:2020tcg, Zioutas:2020cul, Zu:2020idx, An:2020bxd, Choudhury:2020xui, Bramante:2020zos, Jho:2020sku, Nakayama:2020ikz, Primulando:2020rdk, Cao:2020bwd, Paz:2020pbc, Choi:2020udy, Dey:2020sai, Chen:2020gcl, Bell:2020bes, Du:2020ybt, Harigaya:2020ckz, Su:2020zny, Fornal:2020npv, Alonso-Alvarez:2020cdv, Kannike:2020agf, Athron:2020maw, Han:2020dwo, Ema:2020fit, Cao:2020oxq, Kim:2020aua, Borah:2020jzi}, axion-like particles (ALPs)~\cite{Takahashi:2020uio, Long:2020uyf, Li:2020naa, Cacciapaglia:2020kbf, Dent:2020jhf, Bloch:2020uzh, Takahashi:2020bpq, Chiang:2020hgb, He:2020wjs}, solar axions~\cite{Gao:2020wer, Budnik:2020nwz, Coloma:2020voz},\etc

\noindent These experiments are also sensitive to the background from neutrino-electron scattering within the detector~\cite{Ge:2020jfn, Khan:2020vaf, Buch:2020mrg, AristizabalSierra:2020edu, Bally:2020yid, Shoemaker:2020kji}. Such background can originate from solar~\cite{Robertson:2012ib}, atmospheric~\cite{Robertson:2012ib}, supernova~\cite{Beacom:2010kk}, or even reactor neutrinos~\cite{Gelmini:2018ogy}. With increasing sensitivity in the DD experiments, the prospects of discovering anomalous effects on the neutrino floor have become substantial. 
The threshold of the electron recoil energy in the DD experiments are lower compared to the neutrino-electron scattering experiments, such as Borexino, GEMMA,\etc Thus the DD experiments can constrain certain neutrino-related new physics scenarios better than even the dedicated neutrino-electron scattering experiments~\cite{Harnik:2012ni}.   
 The possibility of constraining neutrino interactions in direct detection experiments~\cite{Cerdeno:2016sfi,Harnik:2012ni} and in the  collider searches for DM~\cite{Friedland:2011za,Franzosi:2015wha,Pandey:2019apj} have been discussed in the literature.  

\noindent In this paper, we investigate the constraints on various neutrino NSI originating from a  new light vector boson $Z'$ which kinetically mixes with photons, leading to neutrino-electron scattering. Such a light $Z'$ is often realised as the gauge mediator corresponding to groups, such as $U(1)_{L_{\m}}, U(1)_{L_{\tau}}, U(1)_{L_{\m}- L_{\tau}}$,\etc  But here we do not adhere to any of these specific origins for the $Z'$. Various aspects of such a $Z'$ in connection to NSI have been discussed in the literature~\cite{Farzan:2015hkd, Farzan:2016wym}. Moreover, we  consider several interactions which facilitate active and sterile neutrino conversion and also, interactions with sterile neutrinos altogether. Sterile neutrinos appear in many BSM theories which aim to address the issue of dark matter, neutrino mass and baryon asymmetry in the universe. 
Interactions involving the sterile neutrinos can also possibly explain the excess of events observed at LSND and MiniBooNE~\cite{Bertuzzo:2018itn,Magill:2018jla}. There are several astrophysical and cosmological implications of the interactions considered here, such as stellar cooling, bounds from BBN,\etc In particular, here we consider the bounds from the energy loss of the red giants (RG).  

\noindent In the next section we discuss the NSI under consideration: the existing bounds and the constraints from XENON1T, followed by the concluding remarks.

\section{NSI under consideration: \\Existing constraints and XENON1T}
\noindent {\bf 1a.} \label{1a}
As mentioned earlier, the renormalisable interactions of a light $Z'$ with the neutrinos appear in several well motivated BSM scenarios, such as the aforementioned $U(1)_{L_{\m}}, U(1)_{L_{\tau}}, U(1)_{L_{\m}- L_{\tau}}$,\etc. The interaction is written as
\bea
\mathcal{L} &\supset & g \bar{\nu} \g_{\m} \n Z^{\prime \m} \, .
\label{1a}
\eea 
As mentioned earlier, we consider kinetic mixing, $\mathcal{L} \supset -(\epsilon/4)F_{\m\n}Z^{'\m\n}$. This leads to an additional factor of $\epsilon (q^2 - q_{\m} q_{\n})/m_{Z'}^2$ in the amplitude of neutrino-electron scattering processes. The differential cross-section for neutrino-electron scattering {\it via} this interaction is
\bea 
\frac{d \sigma}{d E_R} = \frac{\alpha g^2 \epsilon^2 m_e}{
 4 (2 E_R m_e + m_{Z'}^2)^2} \Big[ 1 + \Big(1 - \frac{E_R}{E_{\nu}}\Big)^2 - \frac{m_e E_R}{E_{\nu}^2}\Big] , \nn
\eea
where $E_R$, $E_{\nu}$ are the electron recoil energy and the energy of incoming neutrino respectively. We consider the solar neutrino flux with different channels dominant at diffenent energies with the $hep$ process producing neutrinos of maximum energy of 18.7~MeV~\cite{Haxton:2000xb, Bahcall:2004mz}.

\noindent To compute the event rate at XENON1T in presence of this interaction, we use the efficiency factor $\epsilon_{eff}$ presented in ref.~\cite{Aprile:2020tmw}. Under the free energy approximation, 
\bea 
\Big(\frac{d \sigma}{d E_R}\Big)_{tot} = \sum_{i=1}^{54} \Theta(E_R - B_i)\frac{d \sigma}{d E_R},
\eea   
where $B_i$ are the binding energies of the electrons within a Xe atom. The rate of events at XENON1T is given by
\bea
\frac{d N_{R}(E_R)}{d E_{R}}= N_{T} ~\epsilon_{eff}(E_R) \int_{E_{\nu_{\rm{min}}}}^{\infty} dE_{\nu} \frac{d\phi}{dE_{\nu}} \Big(\frac{d \sigma}{d E_R}\Big)_{tot},\nn
\eea
where $E_{\nu_{\rm{min}}}= (E_R + \sqrt{E_{R}^2 + 2 m_e E_R})/2$ and $ N_{T} $ is the density of atoms in the target, which comes out to be 4.2$\times 10^{27}$~ton$^{-1}$ for Xe. We have used $\chi^2$-fitting to find the best-fit points in the presence of NSI. For $m_{Z'}$ less than a few keVs the 2$\sigma$ bound from XENON1T reads $g \epsilon \lesssim 2.4 \times 10^{-13}$ which is slightly stronger than Borexino bound~\cite{Boehm:2020ltd}.

\noindent Stellar dynamics of the sun, red giants, supernova, horizontal branch stars, white dwarfs, {\it etc.} can constrain the light bosons as these stars can abundantly produce such bosons,   leading to anomalous cooling of these stars. In the keV mass range of the $Z'$, the most stringent constraint comes from the cooling of the RGs~\cite{Raffelt}. The typical plasma frequency of such stars is around 8.3 keV and due to the photon-$Z'$ mixing, the plasmons can decay to neutrinos with a decay width of 
\bea
\Gamma (\gamma^{*} \ra \nu \bar{\nu})= \frac{\pi}{12 \omega} \frac{K^2 g^2 \epsilon^2}{(m_{Z'}^2/ K^2 -1)^2}, 
\eea
where $\omega$ is the energy of the photon and $K^2=  \omega_{p}^2$, where $\omega_{p}$ is plasma frequency. The ratio of the decay width $\Gamma_{\n} (\gamma^{*} \ra \nu \bar{\nu})$ for millicharged neutrinos~\cite{Redondo:2013lna}, to that in our case is
\bea
\frac{\Gamma_{\nu}}{\Gamma}= \frac{4 \epsilon_{\nu}^2 e^2}{g^2 \epsilon^2} \Big( \frac{m_{Z'}^2}{K^2} -1 \Big)^{2}.
\eea
The bound on the charge of neutrinos $\epsilon_{\nu}  \lesssim 2 \times 10^{-14} $~\cite{Redondo:2013lna} translates to the constraint $g \epsilon \lesssim 4 \times 10^{-14}$  in our case for $m_{Z'} \lesssim 8.3$~keV. For $m_{Z'} \sim 8.3$~keV, a resonant production of $Z'$ takes place which leads to a comparatively more  stringent constraint. There might be a thermal broadening of this resonant production which has not been considered in this paper.  For $m_{Z'} \gtrsim 8.3$~keV, the plasmon decay width  is suppressed by a factor of $1/m_{Z'}^4$ and both the Borexino and XENON1T bounds surpass the RG cooling bound.

\begin{figure}[h!]
 \begin{center}
 \includegraphics[width=2.7in,height=2.0in, angle=0]{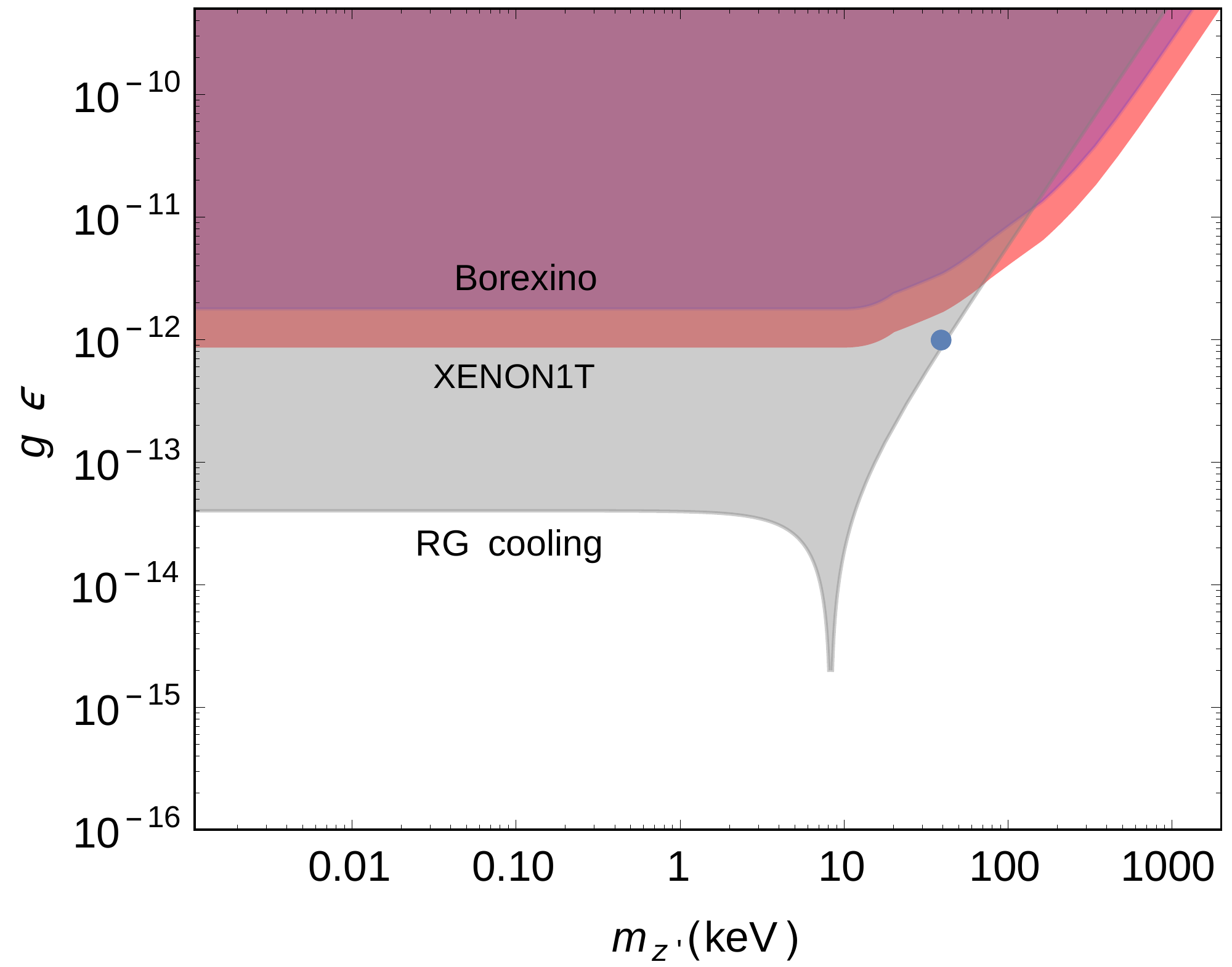}
 \caption{Case {\bf 1a}: The pink, purple and gray regions are ruled out {\it via} XENON1T, Borexino and RG stellar cooling bounds and blue dot is the best-fit point from XENON1T.}
 \label{fig:zpren}
 \end{center}
 \end{figure}

\noindent Along with plasmon decay, the process that contributes to supernova~(SN) cooling is  $e^{+} e^{-} \ra \nu \bar{\n}$. From the observation of Supernova 1987A, the allowed energy loss rate from its core with average temperature $\sim 40$~MeV is $\lesssim 10^{19}$~erg g$^{-1}$ s$^{-1}$, which leads to the upper bound $g \epsilon \lesssim 2 \times 10^{-9}$~\cite{Raffelt}. However, if the coupling exceeds $g \epsilon \sim 2 \times 10^{-7}$ the neutrinos will get trapped inside the core and will not lead to  energy loss~\cite{Davidson:2000hf}. Also such interactions can lead to late decoupling of the neutrinos, leading to the BBN bound $g\epsilon \lesssim 4.2 \times 10^{-9}$~\cite{Davidson:2000hf}. These constraints are much weaker than the RG cooling bounds shown in the fig.~\ref{fig:zpren}.

\begin{figure}[h!]
 \begin{center}
 \includegraphics[width=2.7in,height=2.0in, angle=0]{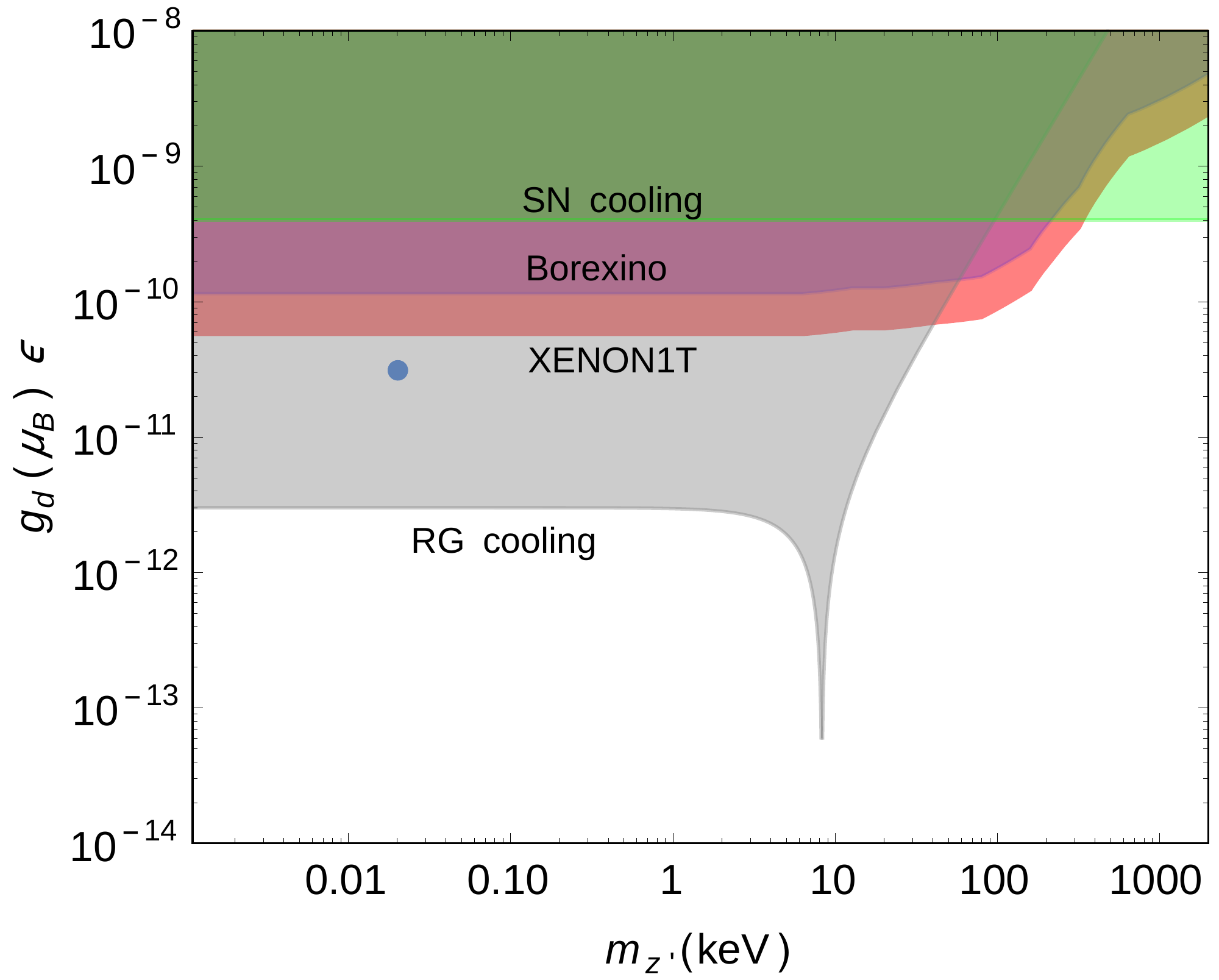}
 \caption{Case {\bf 1b}: The green region is ruled out from SN cooling constraints. Rest of the colour coding are the same as in fig.~\ref{fig:zpren}.}
 \label{fig:1b}
 \end{center}
 \end{figure}

\vspace{15pt}

\noindent {\bf 1b.} \label{1b}
 The neutrino dipole moment with the new $Z'$ boson can also lead to a visible signal at XENON1T. The interaction is defined as
\bea 
\mathcal{L} &\supset & g_d\, \bar{\nu} \sigma_{\m\n} \n Z^{\prime \m\n} \, .
\label{1b}
\eea
This was previously investigated in the paper reporting the possible excess by the XENON1T collaboration~\cite{Aprile:2020tmw}. We revisit the scenario here, presenting a comparison with the bounds from Borexino and RG cooling. The differential cross-section for neutrino-electron scattering with this interaction is
\bea 
\frac{d \sigma}{d E_R} = \frac{\alpha g_{d}^2 \epsilon^2 m_e^2 E_{R}(E_{\nu}-E_{R})}{ E_{\nu} (2 E_R m_e + m_{Z'}^2)^2},  \nn
\eea
where $\alpha$ is the fine structure constant. For a massless $Z^{\prime}$, the preferred range of coupling is $g_{d} \epsilon  \in  (2.8, 5.8) \times 10^{-11} \mu_B$ at 90$\%$ CL, which agrees with the bounds presented in ref.~\cite{Aprile:2020tmw}.
These values of neutrino magnetic moment are allowed from the present Borexino measurement~\cite{Borexino:2017fbd}.
But the constraints on RG cooling render such
values of couplings disfavoured.

\noindent The plasmon decay width in presence of the interaction in eq.~\eqref{1b} is given by
\bea
\Gamma_\text{dip} (\gamma^{*} \ra \nu \bar{\nu})= \frac{g_{d}^2 \epsilon^2}{96 \pi \omega} \frac{K^4}{(m_{Z'}^2/ K^2 -1)^2}.
\eea
As a result, for $m_{Z'} < 8$~keV, the constraint on neutrino dipole moment from RG cooling comes out to be $g_{d} \epsilon \lesssim 3 \times 10^{-12} \mu_{B}$~\cite{Capozzi:2020cbu}.
 Similar to the renormalisable interaction, for $m_{Z'} \gtrsim  10$~keV, the decay width of $Z'$ is suppressed by a factor of $1/m_{Z'}^4$ and the RG cooling constraint becomes weaker than that coming from XENON1T.  However, it should be noted that such masses of $Z'$ will not lead to any excess observed at the detector at $E_R \sim 2$~keV. 
The supernova constraint on this interaction reads $g_{d} \epsilon \in  (4, 200) \times 10^{-10} \mu_B$ and the BBN constraint is $g_{d} \epsilon \lesssim 2 \times 10^{-7} \mu_B$~\cite{Chu:2019rok}.
All the relevant constraints in this scenario are shown in fig.~\ref{fig:1b}.

\vspace{15pt}

\begin{figure}[h!]
 \begin{center}
 \includegraphics[width=2.7in,height=2.0in, angle=0]{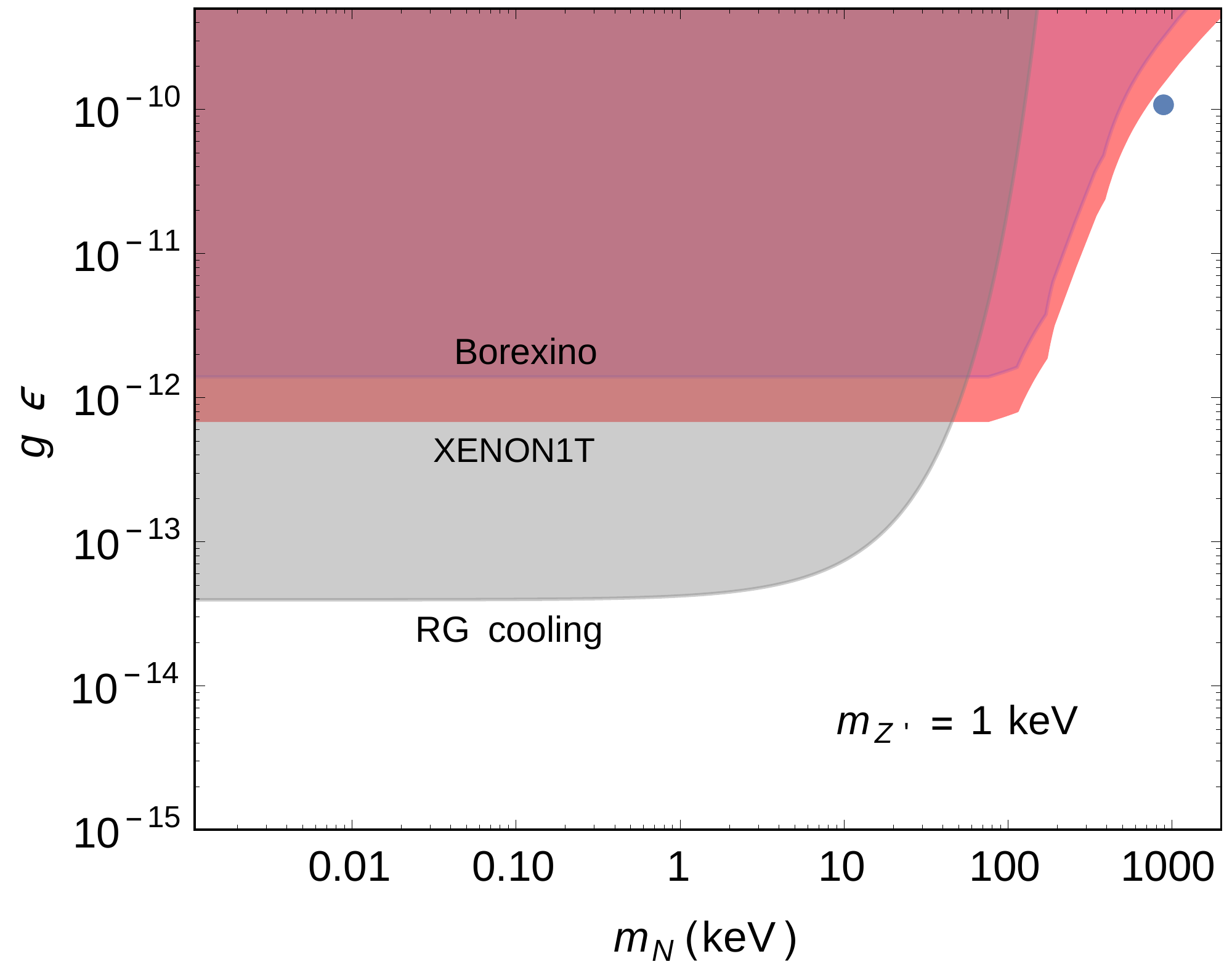}
 \caption{Constraints on the case {\bf 2a} for $m_{Z'}=1$~keV.}
 \label{fig:2a1keV}
 \end{center}
 \end{figure}

\begin{figure}[h!]
 \begin{center}
 \includegraphics[width=2.7in,height=2.0in, angle=0]{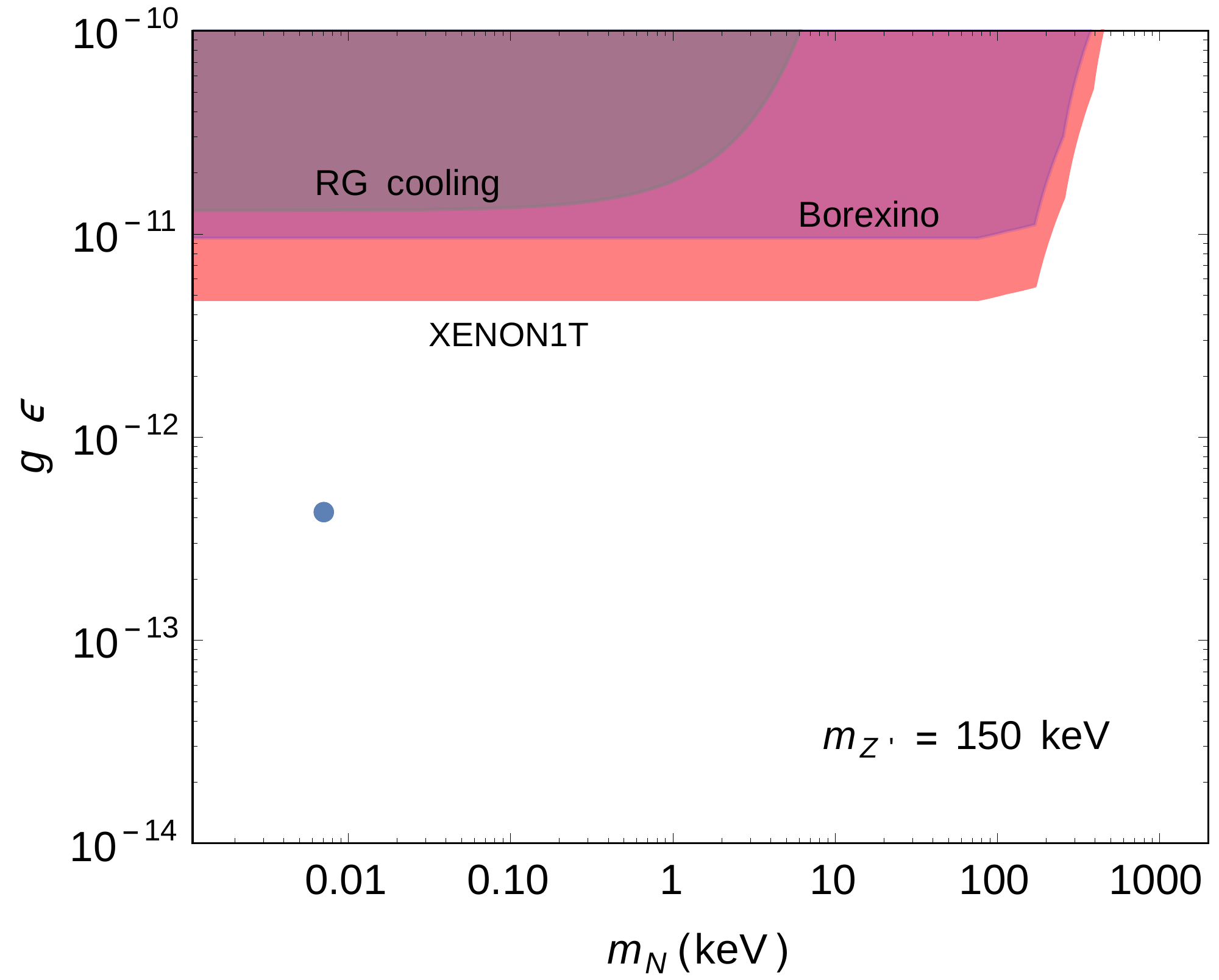}
 \caption{Constraints on the case {\bf 2a} for $m_{Z'}=150$~keV.}
 \label{fig:2a150keV}
 \end{center}
 \end{figure}

\begin{figure}[h!]
 \begin{center}
 \includegraphics[width=2.7in,height=2.0in, angle=0]{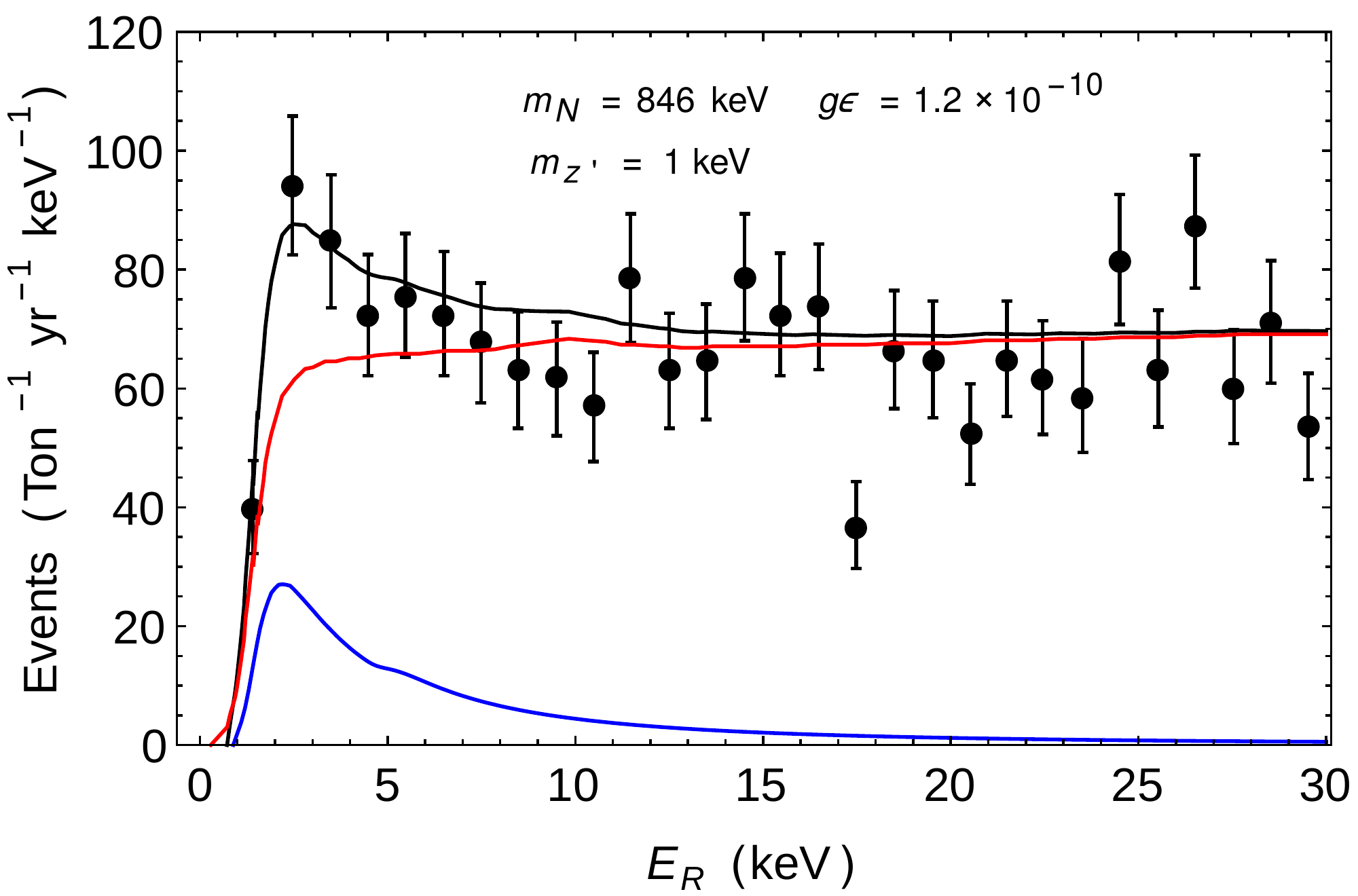}
 \caption{Events at XENON1T with the best-fit point for case {\bf 2a} from fig.~\ref{fig:2a1keV}. The red and blue lines correspond to the background events and the anomalous contribution due to NSI respectively. The black line corresponds to the total number of events in presence of NSI.}
 \label{fig:events}
 \end{center}
 \end{figure}

\noindent {\bf 2a.} \label{2a} Now we consider the renormalisable $Z'$ interaction consisting of an active and a sterile neutrino 
\bea
\mathcal{L} &\supset & g \bar{\nu} \g_{\m} N Z^{\prime \m} \, .
\label{2a}
\eea 
The differential cross-section for $\nu e \ra Ne$ in this case is
\bea
\frac{d \sigma}{d E_R} = \frac{\alpha g^2 \epsilon^2}{4 E_{\nu}^2 (2 E_{R} m_e + 
 m_{Z'}^2)^2}  \Big[4 m_{e} E_{\nu}^2 
 -  m_e m_{N}^2 \nn\\ 
-2 E_{R}(2 m_{e}^2-m_{N}^2-2 m_e E_{R})- 2 E_{\nu}(m_{N}^2+2 m_e E_R)\Big], \nn
\eea
with the minimum value of incoming neutrino energy given as
\bea 
E_{\nu_{\rm{min}}}= \frac{m_{N}^2+2 m_e E_R}{2 \sqrt{E_{R}^2 + 2 m_e E_R} -2 E_R)}. \nn
\eea
As the mass of sterile neutrino $m_N$ increases, the cross-section decreases, leading to a weaker bound by XENON1T.
Coming to the stellar cooling constraints in this case, the production of particles above the plasma frequency will be highly suppressed, thus the stellar cooling is scaled by a factor of $\exp(-m_{N}/\omega_{p})$.  
We have shown the relevant constraints in this scenario on the $m_N - g_{d} \epsilon$ plane in figs.~\ref{fig:2a1keV} and \ref{fig:2a150keV} for $m_{Z'} = 1$~keV and 150~keV respectively.
 The bounds from RG cooling are stronger by around 2-3 orders of magnitude in the former case. As shown in fig.~\ref{fig:events}, this interaction can lead to an excess around $E_R \sim$ 2 keV for $m_{Z'} = 1$~keV at the best-fit point: $m_N =864$~keV, $g \epsilon=1.2 \times 10^{-10}$, which is allowed by the stellar cooling constraints.
Note that, we only consider the sub-MeV sterile neutrinos, whose masses are much lower than the plasma temperature at BBN and inside a supernova. 
Hence, the BBN and SN cooling bounds in this case are the same as in case~{\bf 1a}. 
 
\pagebreak

\begin{figure}[h!]
 \begin{center}
 \includegraphics[width=2.7in,height=2.0in, angle=0]{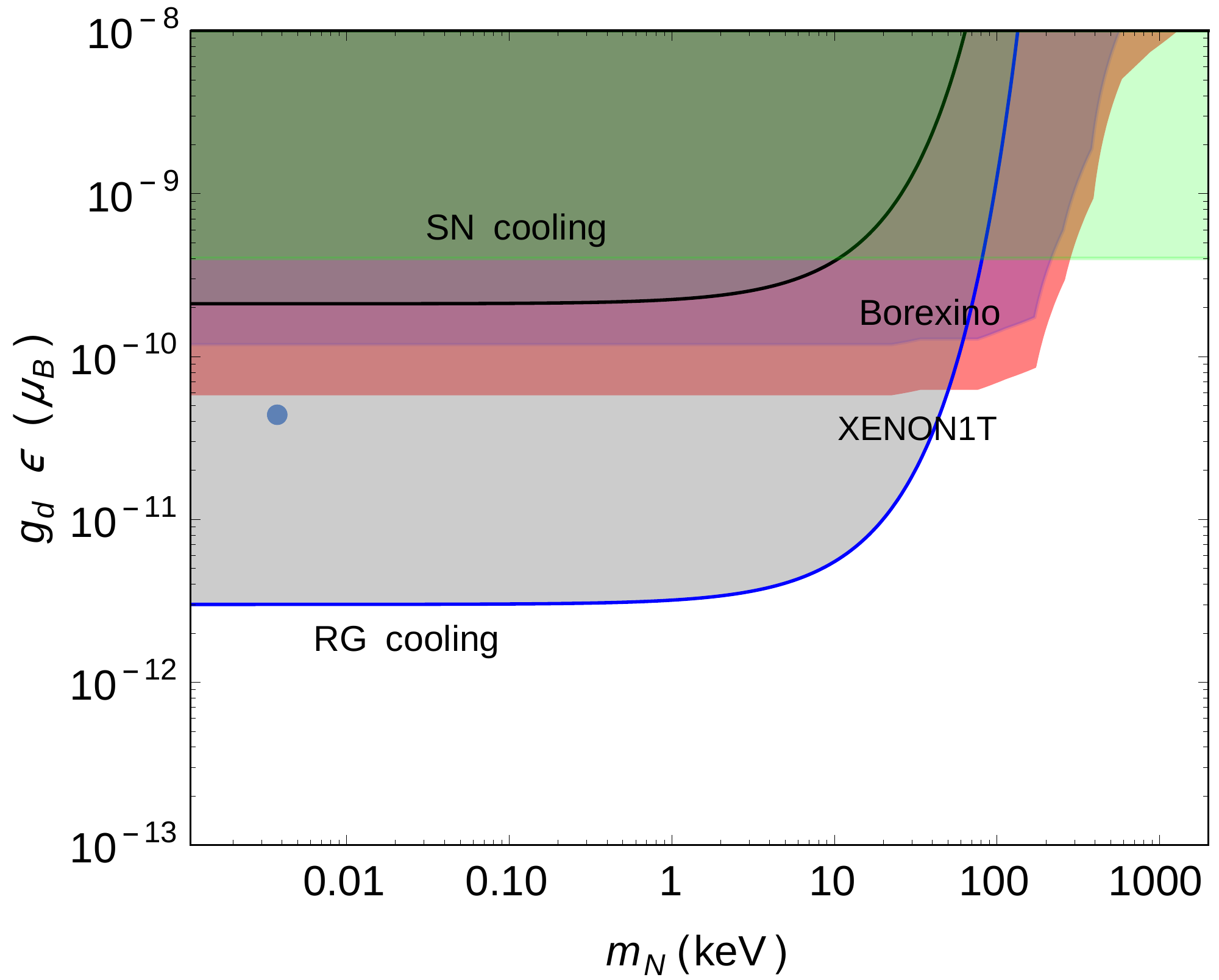}
 \caption{Case {\bf 2b}: The area enclosed by the blue (black) line is ruled out from RG cooling for $m_{Z'}=1$~(70)~keV, whereas the XENON1T and Borexino bounds are similar for both values of $m_{Z'}$. Rest of the colour codings are the same as in fig.~\ref{fig:1b}.}
 \label{fig:2b1keV}
 \end{center}
 \end{figure}

\noindent {\bf 2b.} \label{2b}
The active-sterile neutrino  coupled with $Z'$ {\it via} a dipole term is given as
\bea 
\mathcal{L} &\supset & g_d\, \bar{\nu} \sigma_{\m\n} N Z^{\prime \m\n} \, .
\label{2b}
\eea
This interaction has been studied in light of the reactor anomalies~\cite{Gninenko:2009ks, McKeen:2010rx} . The differential cross-section for the  scattering $\nu e \ra Ne$ in this case is given by
\bea
\frac{d \sigma}{d E_R} &=& \frac{\alpha g_{d}^2 \epsilon^2 m_e^2}{(2 E_R m_e + m_{Z'}^2)^2} \Big[ E_R -\frac{E_R^2}{E_{\nu}} - 
\frac{m_{N}^2 E_{R}}{2 E_{\nu} m_{e}}\nn\\
&& \Big(1- \frac{E_R}{2 E_{\nu}} + \frac{m_e}{2 E_{\nu}}\Big) + \frac{m_{N}^4(E_R - m_e)}{8 E_{\nu}^2 E_{R}^2 m_{e}^2} \Big],\nn
\eea
which is in agreement with ref.~\cite{Shoemaker:2020kji}. We have shown the constraints under consideration on the $m_N - g_{d} \epsilon$ plane in figs.~\ref{fig:2b1keV} for $m_{Z'} = 1$~keV and $70$~keV respectively.
For $m_{N} \lesssim 100$~keV, the constraints from Borexino and XENON1T are of the order $\sim 10^{-10} - 10^{-11} \mu_B$. Here, the constraint from supernova cooling $g_{d} \epsilon \in  (4, 200) \times 10^{-10} \mu_B$ are relevant. 
As mentioned earlier, for sub-MeV sterile neutrinos, the BBN and SN cooling constraints in this case are same as in case {\bf 1b}.

\vspace{5pt}

\begin{figure}[h!]
 \begin{center}
 \includegraphics[width=2.7in,height=2.0in, angle=0]{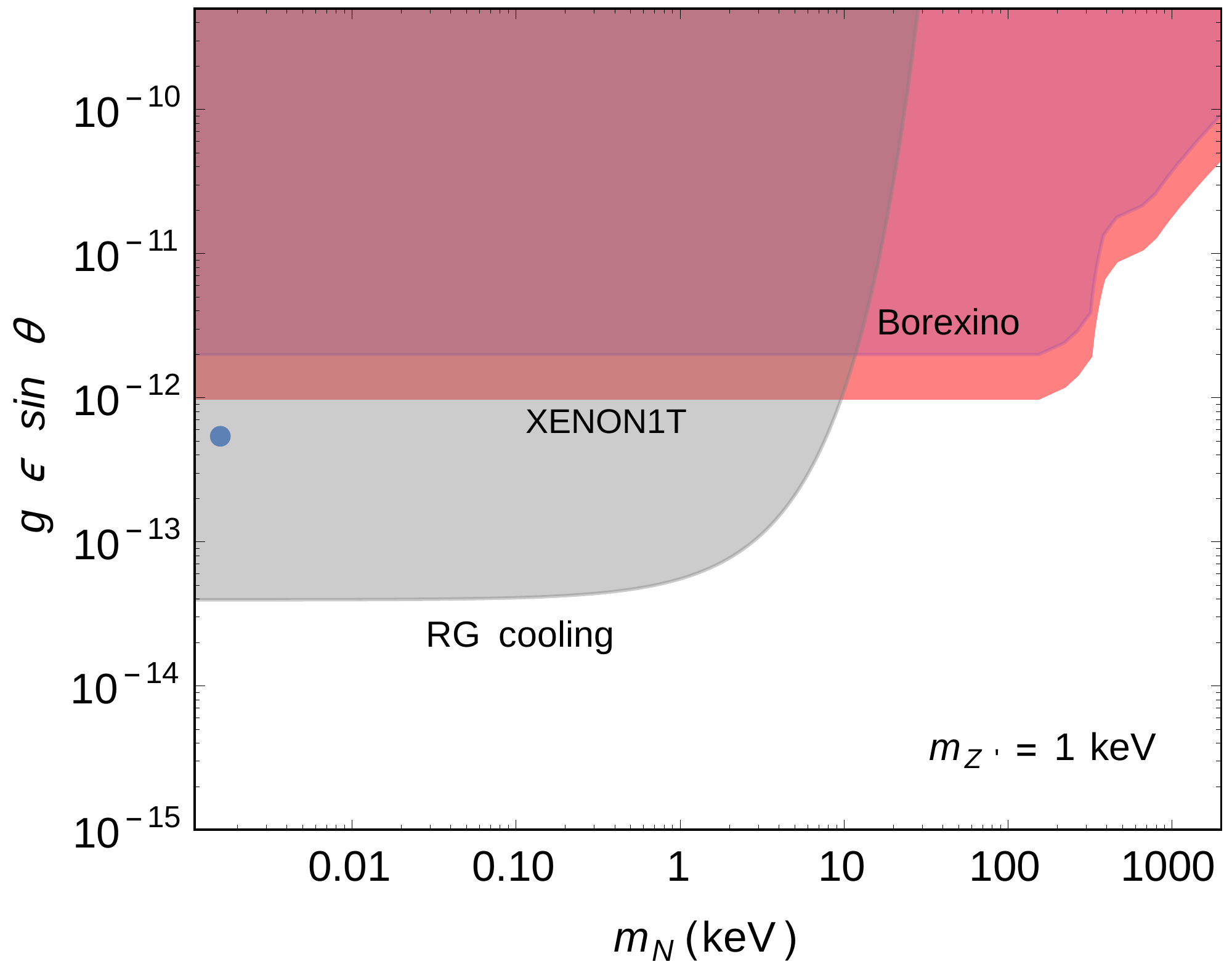}
 \caption{Constraints on the case {\bf 3a} for $m_{Z'}=1$~keV.}
 \label{fig:3a1keV}
 \end{center}
 \end{figure}

\begin{figure}[h!]
 \begin{center}
 \includegraphics[width=2.7in,height=2.0in, angle=0]{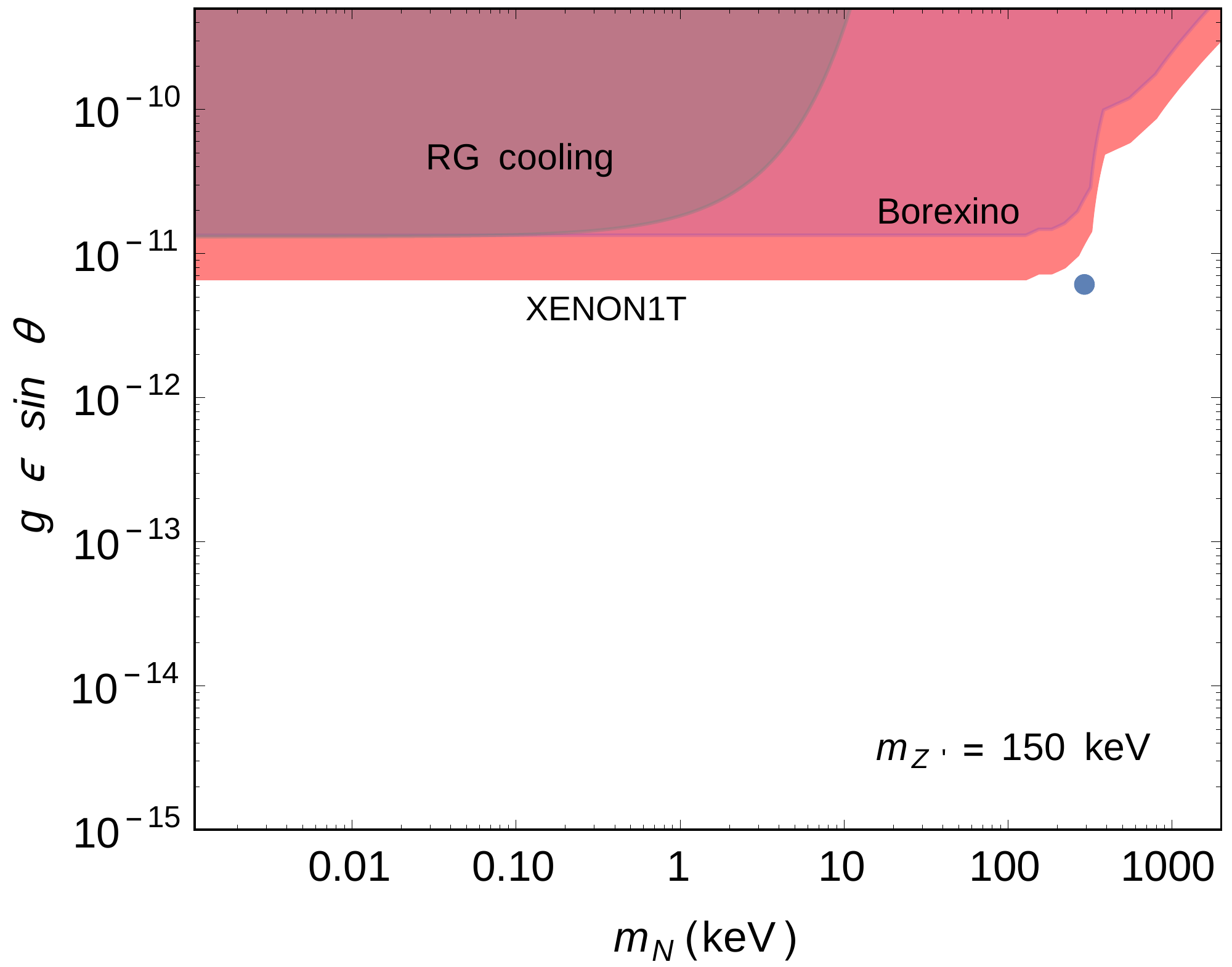}
 \caption{Constraints on the case {\bf 3a} for $m_{Z'}=150$~keV.}
 \label{fig:3a150keV}
 \end{center}
 \end{figure}

\noindent {\bf 3a.} \label{3a} Now we consider the renormalisable  interaction of the $Z'$ with sterile neutrinos which mix with the active neutrinos with an angle $\theta$.   The incoming active solar neutrinos can oscillate to sterile neutrinos~($N$) as they traverse towards the detector and eventually scatter off electrons. 
\bea
\mathcal{L} &\supset & g \bar{N} \g_{\m} N Z^{\prime \m} \, .
\label{3a}
\eea 
The differential cross-section of the scattering $N e \ra Ne$ for such an interaction is
\bea
\frac{d \sigma}{d E_R} &=& \frac{\alpha g^2 \epsilon^2 \sin^2 \theta}{4 E_{\nu}^2 (2 E_R m_e + 
 m_{Z'}^2)^2}  \Big[2 m_{e} E_{\nu}^2 
 - 2 m_e E_{\nu} E_{R} \nn\\
 && -E_{R}(m_{e}^2+m_{N}^2-m_e E_{R})\Big]. \nn
\eea
The minimum energy of incoming neutrino that leads to electron recoil is 
\bea E_{\nu_{\rm{min}}}= \frac{m_e E_R + \sqrt{m_e (E_{R}+ 2 m_e) (2 m_{N}^2+m_e E_R)}}{2 m_e}. \nn
\eea
The constraints on this interaction has been shown in figs.~\ref{fig:3a1keV} and \ref{fig:3a150keV} for $m_{Z'} = 1$~keV and 150~keV respectively. Similar to the previous cases, it can be seen that the best-fit point for $m_{Z'} = 1$~keV is ruled out from the RG cooling constraint, although the same for $m_{Z'} = 150$~keV is still allowed. The neutrino mixing angle $\theta$ can be constrained by the appearance and disappearance experiments. For keV sterile neutrinos these constraints read: $\sin^2 2\theta \lesssim 10^{-2}$~\cite{Diaz:2019fwt, Tanabashi:2018oca}. But such constraints are quite weak compared to the XENON1T bounds presented here. Moreover, if the sterile neutrino is realised as a DM candidate, its stability has to be ensured, because it can decay through $N \rightarrow \nu \gamma$. If the heavier generations~($N_{2,3}$) of $N$ are also present, those can subsequently decay as $N_{2,3}  \rightarrow N \gamma$, leaving imprints on the CMB observables. Because of the mixing of the sterile neutrino with active neutrinos, it can contribute to the total lepton asymmetry in a leptogenesis scenario which, in turn, provides a bound on $\sin \theta$~\cite{Boyarsky:2009ix}.

\noindent In the limit where the neutrinos in the final state are massless, the decay width for $\nu_i \ra \nu_j \nu_k \bar{\nu}_l$ is given by 
\bea
\Gamma_i= \frac{g^4 |U_{Ni} U_{Nj} U_{Nk} U_{Nl} |^2}{192 \pi^3} \Big( \frac{m_{i}^5}{m_{Z'}^4} \frac{m_i}{E_{\nu}}  \Big),\nn
\eea    
where $U$ is the mixing matrix and the factor $m_i/E_{\nu}$ takes care of the time dilation. If $|U_{Ni}| \sim 0.1$ for active-sterile mass states, $|U_{N4}| \sim 1$, $m_4 \sim m_N = 1$~keV and $m_{Z'} \sim 150$~keV, the lifetime of the sterile neutrinos comes out to be $\tau=\Gamma_i^{-1}=1.5 \times 10^{30}\, (8 \times 10^{-5}/ g)^4 (E_{\nu}/ 10~\text{MeV})$~cm. Thus the effect of $N$ decay cannot be observed for solar or atmospheric neutrinos.  Also, $g \gtrsim 8 \times 10^{-5} $ allows $N$ to completely decay into active neutrinos before recombination, thereby evading the bounds from neutrino mass~\cite{Denton:2018dqq}.   

\vspace{40pt}

\vspace{30pt}

\pagebreak

\begin{figure}[h!]
\begin{center}
\includegraphics[width=2.7in,height=2.0in, angle=0]{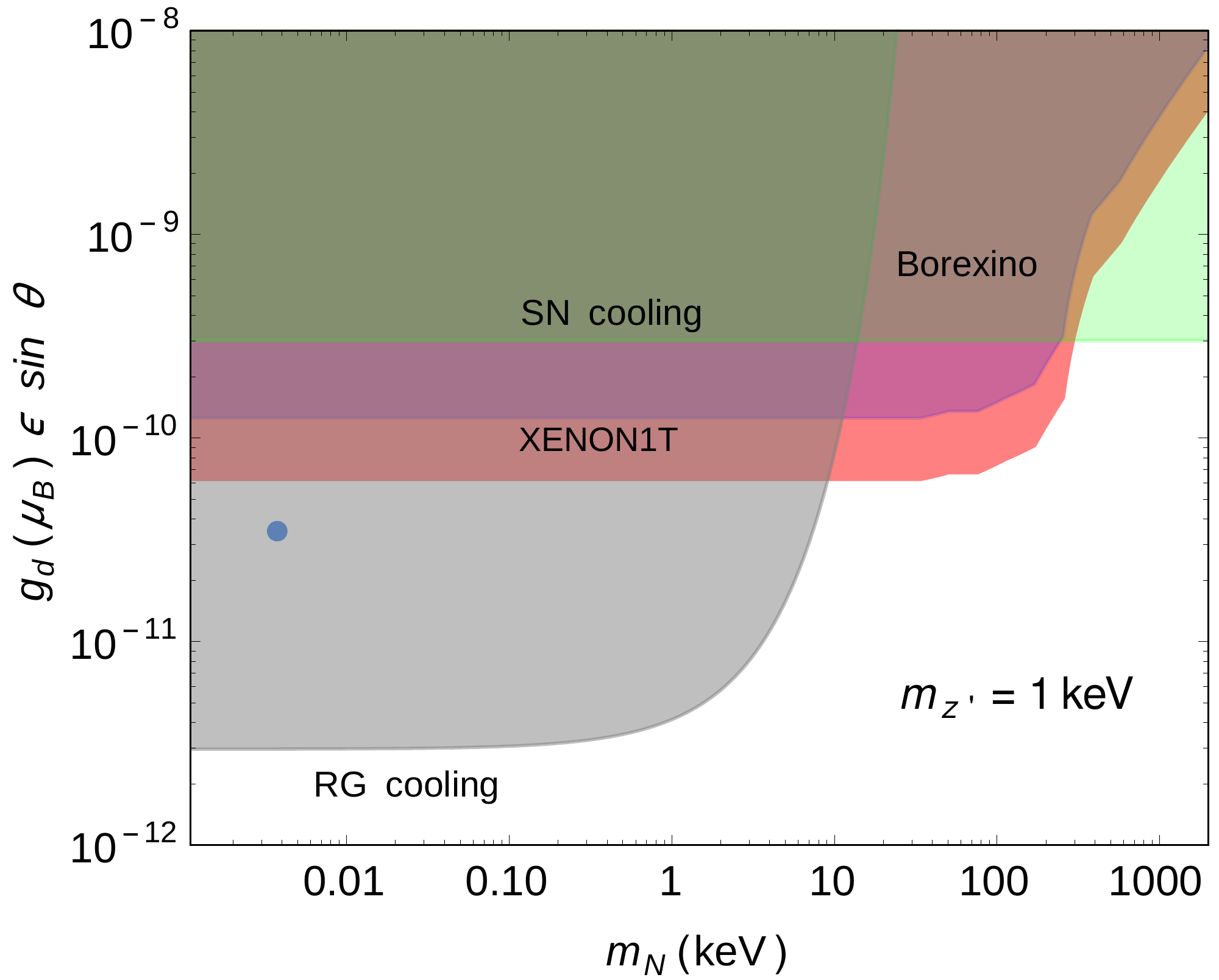}
 \caption{Constraints on the case {\bf 3b} for $m_{Z'}=1$~keV.}
 \label{fig:3b1keV}
 \end{center}
 \end{figure}

\begin{figure}[h!]
 \begin{center}
 \includegraphics[width=2.7in,height=2.0in, angle=0]{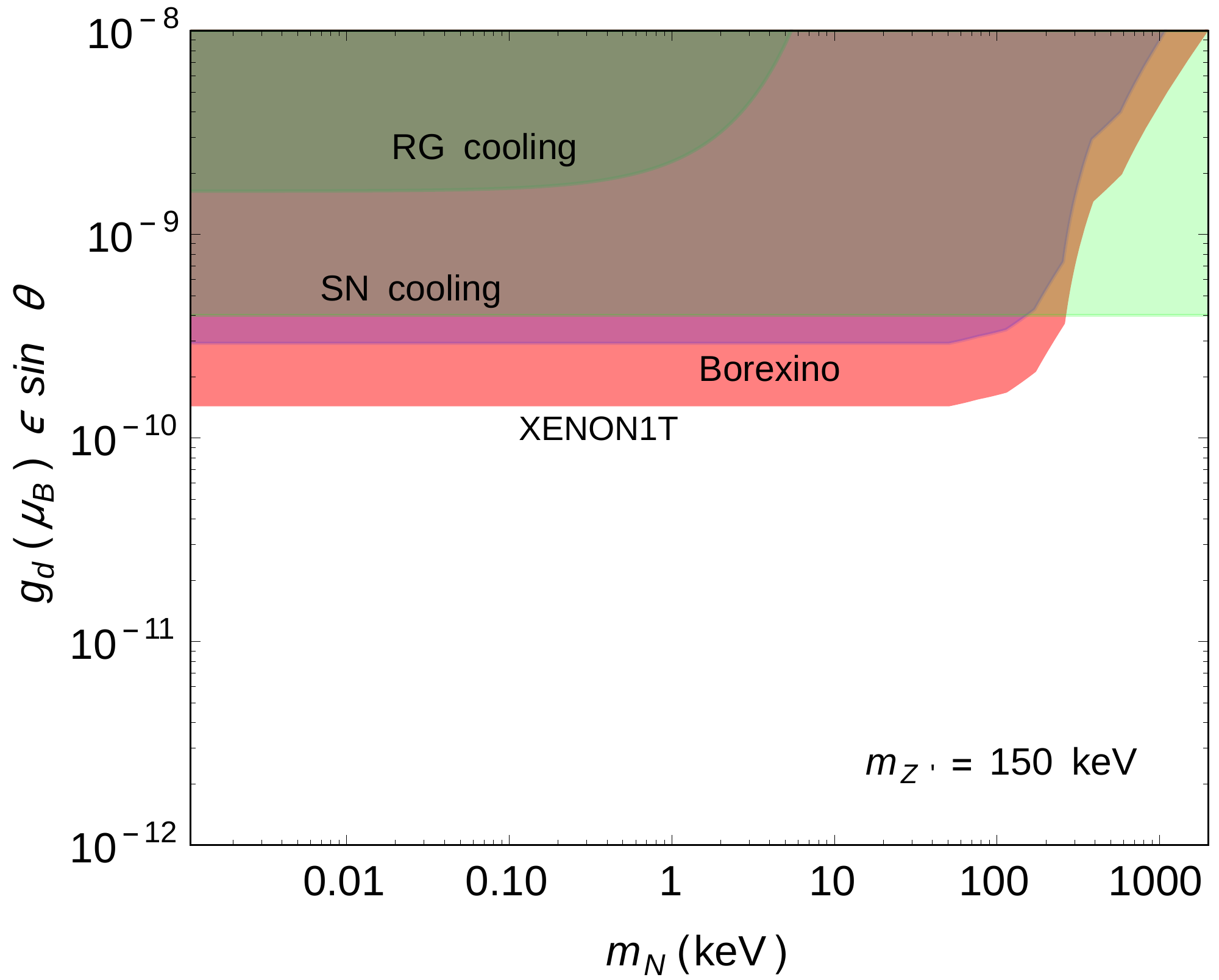}
 \caption{Constraints on the case {\bf 3b} for $m_{Z'}=150$~keV.}
 \label{fig:3b150keV}
 \end{center}
 \end{figure}

\noindent {\bf 3b.} \label{3b}
If the active and sterile neutrinos mix with an angle $\theta$ and the new vector boson $Z'$ interacts with the sterile neutrinos {\it via} a dipole term, the Lagrangian is given as
\bea 
\mathcal{L} &\supset & g_d\, \bar{N} \sigma_{\m\n} N Z^{\prime \m\n} \, .
\label{3b}
\eea 
The differential cross-section for sterile neutrino-electron scattering  in this case is
\bea
\frac{d \sigma}{d E_R} &=& \frac{\alpha \sin^2 \theta g_d^2 \epsilon^2 m_e E_R}{(2 E_R m_e + m_{Z'}^2)^2} \nn\\
&&\times \Big[m_{e} +  
\frac{m_{N}^2}{2 E_{\nu}^2}  
(E_R- 2 m_e) - \frac{m_e E_R}{ E_{\nu}} \Big]. 
\label{NNdip}
\eea
\noindent The constraints on this interaction are shown in figs.~\ref{fig:3b1keV} and \ref{fig:3b150keV} for $m_{Z'} = 1$~keV and 150~keV respectively.  
 The maximum contribution to the  solar neutrinos flux stems from the $pp$ and $^7Be$ channels with the maximum energy around 400~keV, after which the flux decreases. Thus the XENON1T constraints get weaker above $m_{N} \sim$ 200~keV. Also, in the limit $m_{N} \sim E_{\nu}$ the differential cross-section for $N e \ra N e$ comes out to be greater than $\nu e \ra N e$ in case {\bf 2b}. Thus, for $m_N \gtrsim 400$~keV, the constraints in this case are stronger than the case {\bf 2b}.

\noindent The BBN constraints on sterile neutrinos can be evaded by invoking self-interaction between the active and sterile neutrinos which leads to an effective potential. Thus, the mixing angle between active and sterile neutrinos are suppressed at high temperatures, thereby reducing the production of sterile neutrinos~\cite{Dasgupta:2013zpn}. Similar to the previous interaction, the cosmological bounds on the sum of the neutrino masses can be evaded if the sterile neutrinos decay to active neutrinos, which is facilitated by this interaction~\cite{Denton:2018dqq}.

\noindent Along with the dipole terms, another dim-5 NSI leading to neutrino-electron scattering can be written as 
\bea
\mathcal{L} \supset  g_{D} (\bar{\nu}_i i \overset\leftrightarrow{\partial^{\m}} \nu_j )  Z'_{\mu}, 
\label{opzp}
\eea
where $\nu_{i,j} = \nu,~N$. 
The differential cross-section with this interaction is similar to the dipole case up to an additional factor of 4. For example, the differential cross-section for $N e \ra N e$ {\it via} the  interaction in eq.~\eqref{opzp} comes out to be four times of that in eq.~\eqref{NNdip}. Hence, all the bounds in this case can be obtained from figs.~\ref{fig:3b1keV} and \ref{fig:3b150keV} with the scaling $g_{D} \epsilon \sin \theta = 2 g_{d} \epsilon \sin \theta$.

\section{Conclusion }
\noindent Increasing sensitivity in the DD experiments make them  potential probes of the neutrino floor as well. The neutrino floor is modified in the presence of non-standard neutrino interactions. Recently, the observations of XENON1T has provided stringent constraints on various new physics scenarios. In this letter, we have presented the XENON1T constraints on neutrino NSI through a kinetically mixed light $Z'$, along with the interactions of $Z'$ with sub-MeV sterile neutrinos. 
 It is expected that such constraints are flavour universal. We  compare these bounds with the constraints from Borexino and red giant cooling.

\noindent For renormalisable interactions with active neutrinos,\ie in case {\bf 1a}, the RG cooling constraints are the strongest for  $m_{Z'} \lesssim 80$~keV, beyond which the XENON1T bound becomes most stringent.
In case {\bf 1b},\ie for dipole interactions of active neutrinos, XENON1T provides the strongest constraints for $30~\text{keV} \lesssim m_{Z'} \lesssim$~300~keV. For $m_{Z'} \lesssim 30$~keV and $m_{Z'} \gtrsim 300$~keV the RG cooling and SN cooling bounds prevail over the XENON1T constraint respectively.

\noindent For NSI involving heavy sterile neutrinos and lighter $Z'$, RG cooling provides the most stringent constraints for $m_{N} \lesssim 10$~keV, above which XENON1T bounds are dominant. But with a heavier mediator, for example, $m_{Z'} \sim $ 150~keV, XENON1T bounds are always stronger than that from RG cooling. 
In the cases {\bf 2b} and {\bf 3b},\ie the dipole interactions with sterile neutrinos,  the SN cooling constraints dominate above $m_N \gtrsim 350$~keV. 
The $pp$ and $^7Be$ channels of neutrino production, which are the most dominant components of the solar neutrino flux, fall steeply for $E_{\nu } \gtrsim 400$~keV. Hence, the XENON1T constraints on interactions involving sterile neutrinos get weaker for $m_N \gtrsim 200$~keV.
We have seen that, for $m_N \sim E_\nu$, the process $N e \ra N e$ at XENON1T has a higher cross-section than the inelastic scattering $\nu e \ra N e$, leading to a tighter bounds on $Z'$ interactions with two sterile neutrinos for $m_N \gtrsim 400$~keV. 

\noindent It has been seen that most of the NSI scenarios that can explain the XENON1T excess around $E_R \sim 2$~keV are in tension with the RG cooling bounds. Though, we have pointed out that there could be a scenario, namely case {\bf 2a}, which can  potentially explain the aforementioned excess even after considering the RG cooling bounds.

\hfill

\end{document}